\magnification=\magstep1
\centerline{\bf HIERARCHICAL CLUSTERING AND GALAXY FORMATION}
\medskip
\centerline{\it Simon D.M. White}
\smallskip
\centerline{\it Max-Planck-Institut f\"ur Astrophysik, Garching bei
M\"unchen, Germany}
\medskip
\centerline{Position paper for the 78th Dahlem Conference: The
Evolution of the Universe}
\bigskip

\centerline{\bf ABSTRACT}

\noindent{I review the theory of hierarchical clustering, starting
with an historical
overview and moving on to a discussion of those aspects of dissipationless 
clustering under gravity which are most relevant to galaxy formation. 
I conclude with some comments on the additional problems which arise
when including all the other physics needed to build a realistic
picture for the origin and evolution of the galaxy population.}
\bigskip
\noindent{\bf 1 An historical introduction}
\smallskip

The idea that structure in the Universe might build up through hierarchical 
clustering became popular in the late 1960's and early 1970's primarily
as a result of the work of Jim Peebles and his collaborators. These
developments, and indeed much of the material in this 
introduction, are reviewed from a somewhat different perspective 
in Peebles's own textbooks (1980, 1993) and in the recent text by 
Padmanhaban (1993). Soon after the
discovery of the microwave background had ensured the position of the Hot 
Big Bang as the dominant cosmological model Peebles and Dicke realised that
when the primordial plasma became neutral at a redshift of 1000 its Jeans 
mass would drop from very large values to about $10^6 M_\odot$. This led them
to suggest that a large population of globular cluster-like objects might 
collapse immediately after recombination, and that larger systems might form
subsequently by aggregation of these first objects. Although this specific
hypothesis immediately encountered a number of difficulties, the picture that
small things should collapse first and then merge together to make larger 
objects remained as what became known as the isothermal theory of 
structure formation. 

The name ``isothermal'' originates from a classification of perturbations
in a radiation-dominated universe. By the late 60's it was known that there 
are two independent pertubation modes of the coupled radiation-gas mixture
for which the density contrast of the matter fluctuations is a non-decreasing
function of time. For the isothermal mode the radiation temperature is
almost uniform at early times but the photon-to-baryon ratio varies from place
to place. Such fluctuations survive with little damping until matter and
radiation decouple, at which time overdense regions with mass exceeding the 
matter Jeans mass are able to collapse. Since there was no known physical 
mechanism to generate such fluctuations, it was unclear what assumption to 
make about their relative amplitudes on different scales. The lack of an 
obvious characteristic mass in the range of interest $10^6<M/M_\odot<10^{17}$ 
suggested a power-law fluctuation spectrum, but what power-law
index is appropriate? During the 1970's most work adopted a white-noise 
spectrum corresponding, perhaps, to a Poisson distribution of the first
collapsed objects. While everyone realised that this was an assumption of
convenience (and one that was often challenged) it nevertheless shaped the
prevailing picture of ``generic'' hierarchical clustering.

The second important perturbation mode, known as the adiabatic mode, has 
uniform photon-to-baryon ratio but spatially varying temperature, density
and curvature. This is often considered to be the ``naturally'' dominant
mode since it grows faster than the isothermal mode. Thus if both modes
were stimulated with comparable amplitude in the early universe, the adiabatic 
mode would dominate at late times. In the period leading up to recombination
adiabatic fluctuations are damped strongly by photon diffusion on all scales 
smaller than the Silk mass ($\sim 1.4~10^{15}(\Omega_0h^2)^{1/4}M_\odot$ for
the baryon density inferred from cosmic nucleosynthesis). If such modes are
dominant the first structures to collapse have masses much larger than those
of galaxies. This ``adiabatic'' picture for structure formation was championed
by Zel'dovich and his school who pointed out that the initial collapse would
generically be one-dimensional and would therefore give rise to coherent
sheet-like structures which they termed ``pancakes''. Galaxies would have to
form by the fragmentation of these pancakes.

These two competing pictures set up two views of structure formation which are
still with us today despite the fact that the specific models on which they 
were based are no longer considered viable. In the isothermal world-view 
objects of galaxy scale form by aggregation and merging, while large-scale 
structures are essentially random and have little influence on galaxy 
properties. In the adiabatic world-view large-scale structure displays 
considerable coherence and its collapse dynamics determine where and how 
galaxies form. (Few people would still argue that galaxies form by the 
fragmentation of bigger objects because they appear older than the observed 
large-scale structure.) These different 
points of view were reinforced by the use of different mathematical tools. 
Peebles and his collaborators took their techniques from statistical physics --
correlation functions, the BBGKY hierarchy etc. -- while Zel'dovich and his
group applied results from the theory of Hamiltonian flows -- singularity
classification, topology of structure, and so on. The first approach clearly
emphasises stochastic properties while the second emphasises large-scale
coherence. Good techniques for simulating and visualising either hierarchical 
clustering or coherent collapse from a random field became available only
in the mid 1980's. It is interesting that even today the language used
to analyse and interpret such simulations can often be traced back to one
or other of the original ``schools''.

By the time that large numerical simulations became possible the most popular 
cosmogonies assumed the dominant dark matter component to be some kind of free
elementary particle. The successor of the adiabatic picture was the 
neutrino-dominated or HDM (Hot Dark Matter) model. Simulations of HDM showed 
that evolution from a gaussian random field with a well-defined coherence 
length proceeds quite rapidly from a state where very little matter is in any 
nonlinear object to one where more than 25\% of all matter is in collapsed and 
virialised clusters with mass comparable to the coherence scale
(White et al. 1983). In the 
intermediate regime a connected structure built of highly asymmetric elements 
does indeed form, but the dominant visual impression is of a network of 
filaments rather than of a cellular foam made up of sheets. The successor of 
the isothermal picture was the CDM model. It has the important feature that its
power spectrum is significantly redder than white noise ({\it i.e.} the power
density at high spatial frequencies, corresponding to galaxies, is well below
that on the scale of galaxy clusters). As a result collapse on galaxy scales 
occurs more recently in the CDM universe than envisioned by the older model. 
In addition there is a surprising coherence of structure on scales larger than
galaxies (White et al. 1987). This coherence, again an apparent
network of filaments, is even 
stronger in recent galaxy surveys and in variants of CDM which attempt to fit 
these surveys and to accommodate the fluctuations measured by COBE.

These numerical developments have made the HDM incarnation of the adiabatic
picture seem unattractive while at the same time showing that CDM cosmogonies
are much less clearly hierarchical than the old isothermal picture and
can profitably be analysed using the language of coherent large-scale flows.
These issues are mainly relevant to the topic of this paper because they
mean that the formation of galaxies cannot easily be separated from that of 
larger and smaller objects in CDM-like models. Thus protogalactic collapse
is neither the falling together of a single smooth perturbation nor
the merging of a set of well-equilibrated precursor objects, but lies somewhere
between the two. Similarly, while galaxies generally form before the larger 
structures in which they are embedded, the temporal separation of the two 
processes is not enough for them to be independent. As a result substantial
``biases'' can arise in the galaxy population ({\it i.e.} the properties
of galaxies can end up depending strongly on their large-scale environment).
I come back to both these issues in later sections.

N-body simulations have also clarified another important question about 
hierarchical clustering. Measurements of the two-point correlation function 
for galaxies show a well defined power-law continuing down to scales of a
few tens of kpc where the distribution is highly nonlinear. Furthermore
three-point and higher order correlations are related to the two-point
function in a simple way which appears almost independent of scale. Impressed
by his discovery of these facts in the 1970's Jim Peebles suggested that
the galaxy distribution and the underlying mass distribution might form
some kind of scale-invariant or fractal-like hierarchy, and that the 
continuation of power-law behaviour to very small scales might reflect
the dynamical stability of this arrangement (e.g. Peebles 1978). In 
contrast, Martin Rees and I
argued that a virialised clump of non-dissipative dark matter would
not maintain a hierarchical structure but would evolve into a monolithic
dark halo with a well-defined centre and a smooth density profile
(White \& Rees 1978). We
inferred from this that galaxy clusters must contain many galaxies rather 
than a single ``supergalaxy'' because dissipative processes concentrated
the galaxies sufficiently during formation for them to be able to avoid 
``overmerging''. This issue has remained controversial but most numerical 
studies now agree that objects formed by hierarchical clustering of
dissipationless matter from gaussian initial conditions do not retain much
significant substructure. I will return to these matters later.

The question of overmerging brings us to a critical point. While merging
of dark halos may occur rapidly during hierarchical clustering, it is much 
less clear whether merging is an important process in the formation and
evolution of individual galaxies. Toomre has argued forcefully that elliptical
galaxies form by the merger of disk systems (Toomre 1977). His idea
has gained strong support
from two directions. Simulations by Barnes, Hernquist and others have shown 
that the process does indeed produce objects with the right kind of
structure,
while observers have found real systems in which this transformation is 
currently occurring and have shown the internal structure of ``normal'' 
ellipticals to possess much of the diversity expected in merger
products (Barnes 1995). 
Another line of argument, due principally to Ostriker and his colleagues, notes
that the giant cD galaxies seen at the centres of many rich clusters may result
from galaxy merging during the formation and evolution of the cluster
(e.g. Hausman \& Ostriker1978). While 
the direct observational evidence for this process is not fully convincing, it
seems a natural extension of Toomre's idea since in most of their properties 
cD galaxies form a smooth continuation of the sequence of ordinary bright 
ellipticals. The main remaining questions are whether a merger origin can be 
consistent with the systematic regularities of the elliptical population ({\it
e.g.} the ``fundamental plane''), and if so, then exactly what kind of objects 
merged at which epoch. It seems unlikely that present ellipticals could have 
arisen through the merging of randomly chosen objects from the present disk 
galaxy population, although even this issue remains controversial.

In contrast, it is generally agreed that the stellar disks seen in spiral 
galaxies could {\it not} have formed through the aggregation of pre-existing
stellar systems. Rather the material of the disks must have settled into
its present thin and rotationally supported configuration while 
still gaseous (and hence dissipative) and must have remained relatively 
undisturbed since the bulk of it turned into stars. T\'oth \& Ostriker
(1992) noticed
that this requirement places limits on the rate at which even quite small 
galaxies are merging with present-day spirals. They concluded that an open 
universe is required for the current accretion rate to be 
sufficiently small. Their argument is clearly important enough to merit
further detailed investigation. If spiral disks cannot be made by mergers
then the same might seem to be true for the bulges at their centres despite the
many similarities between bulges and ellipticals. This conclusion need not
apply if mergers were to produce bulges sufficiently early that the disk
could be accreted later. I will argue below that this sequence is
indeed viable in hierarchical clustering, even for a high density universe.
Note, however, that the gas which settles into the disk could not have
been coextensive with the stars in the premerger systems since it would then
produce a disk which is smaller and more strongly bound than the bulge
rather than the opposite which we observe.

The above paragraphs cover only rather general points about how hierarchical
clustering may affect galaxy formation. However, within specific models
for hierarchical clustering it is possible to predict how, when and where 
galaxies form, what they merge with, how their various components are 
differentiated, and what sets the relationships between the distributions of 
galaxies and of mass and between galaxy properties such as morphology and 
luminosity and the larger scale environment. In the remainder of this paper
I will explore the simplest such clustering model, which is  based on 
dissipationless gravitational collapse from an initially gaussian distribution
of density perturbations. As a result of intensive analytic and numerical study
this model is now quite well understood. I will argue that when implemented
in the context of a CDM-like cosmogony it can reproduce most of the 
qualitative and many of the quantitative properties of the observed galaxy 
population. In addition it provides a phenomenology which is very helpful
when interpreting the data now becoming available for high redshift galaxies,
and it suggests how such observations may be used both to test the hierarchical
clustering paradigm and to estimate the parameters of the specific cosmogony 
in which it is implemented.

In section 2 I discuss purely dissipationless hierarchical clustering, I point
out a number of regularities of the gaussian case and of the Press \& Schechter
model for its evolution, I set out what I now believe to be well established 
and what I consider still uncertain, I discuss some new work on the expected
structure of dark halos, and I note the points of contact which can already
be made with observation. In section 3 I consider the new issues that can
be addressed by studying what happens to a dissipative gas component which
clusters with the dark matter. Interesting points arise concerning the
angular momentum of galaxy disks and the relative amounts of gas, stars and
dark matter seen in galaxy clusters. Finally in section 4 I sketch the results
obtained so far by combining these techniques with simple phenomenological 
models for the formation and evolution of the stars in galaxies. Most
of the material in these sections is presented in fuller form in my 
lecture notes for the 1993 Les Houches summer school (White 1996).
\medskip
\noindent{\bf 2 Dissipationless clustering}
\smallskip
\noindent{{\it 2.1 Gaussian or not?~~~}The basic requirements for hierarchical
clustering are that the growth of structure should be driven by gravity and
that small things should collapse first. In this paper I will consider 
models in which the dominant mass component clusters dissipationlessly under
gravity. Thus the dark matter must either be some kind of free elementary
particle, or a population of black holes, stellar remnants, or ``jupiters''
which formed well before the collapse of objects of galactic scale. It may
be that some fraction of the dark matter formed at relatively late times 
through cooling flows or other means. Provided this fraction is not too large
such a complication would not greatly affect my arguments. Small objects will 
form first provided the {\it rms} fluctuation of the initial density within a 
smoothing filter enclosing mass $M$ is a decreasing function of $M$. 
Equivalently $k^3P(k)$ must be an increasing function of spatial wavenumber $k$
where $P(k)$ is the power density in a fourier decomposition of the initial
density field.}

The initial field will be gaussian if and only if the phases of its different 
fourier modes are uniformly and independently distributed. There are certainly
plausible hierarchical clustering models for which this condition is not 
satisfied. Examples of particular interest arise in theories where density 
fluctuations are generated by cosmic strings or textures. At present it is 
still unclear how strongly the behaviour of such models will deviate from that
of a gaussian model with similar $P(k)$. To the extent that the effective 
density fluctuation at a point results from superposing the influence of many 
strings or textures, it seems possible that the Central Limit Theorem may lead
to approximately gaussian behaviour. From now on I will restrict myself
entirely to gaussian models for which $P(k)$ gives a complete description
of the statistical properties of the initial conditions and hence of the
subsequent growth of structure both linear and nonlinear.
\medskip
\noindent{{\it 2.2 How should we describe hierarchical clustering?~~~}The 
easiest way is to begin with the simplest possible case and then to extend it 
to cover more realistic possibilities. Consider a universe containing only 
collisionless matter. Assume that at some very early ``initial'' time
the density field was gaussian with a power-law fluctuation spectrum 
$P(k)\propto k^n$ and particle motions were negligible. At some much later 
time, after the universe has expanded by a factor $a$, the amplitude of those 
fluctuations which are still linear will have increased by a factor $b(a)$ 
where $b(a)=a$ for the simplest case of an Einstein-de Sitter universe. We can
therefore define a characteristic wavenumber $k_*(a)$ which separates linear 
from nonlinear scales by setting $b(a)^2k_*^3P(k_*)=1$. This in turn defines a 
characteristic mass $M_*$ for nonlinear objects where $M_*(a)\propto
b(a)^{6/(3+n)}$.}

In the Einstein-de Sitter case the universe itself expands as a power law
$a\propto t^{2/3}$ and so defines no characteristic time, length or mass.
For power law initial fluctuations it then seems natural to assume that the
growth of structure will be self-similar at late times. This implies that
{\it all} the statistical properties of the structure are independent of
time once masses are expressed in units of $M_*(a)$, lengths in units of 
$a/k_*(a)$ and time in units of the age of the universe. It is important
to note that self-similarity is an {\it assumption} and has not been proven.
In fact, there is some dispute over the values of $n$ for which self-similar
evolution is possible. Hierarchical clustering requires $n>-3$, while $n\leq 4$
is required for any physically plausible fluctuation distribution. However, the
full range $-3<n\leq 4$ may not give rise to self-similar evolution. For
$n\geq 1$ the binding energy of $M_*$ objects is dominated by the internal
binding energy of the smaller objects from which they form and so cannot
scale in the expected way with $a$. In this case self-similar
evolution is possible
only if nonlinear objects do not relax to form monolithic halos but instead
maintain a hierarchical structure down to arbitrarily small scales. The limited
simulation data available do not support this behaviour. For $n\leq -1$ the 
(linear) contribution of large-scale perturbations to the {\it rms} bulk motion
of objects is divergent and $P(k)$ must be cut off below some suitably small 
$k_c$ in order to get a viable model. It seems unlikely that this will affect 
the way nonlinear structures build up while $k_*\gg k_c$, and so I would claim
self-similar clustering to be a plausible hypothesis for $-3<n<1$.

The range of $n$ which seems likely to be relevant for the formation of
nonlinear objects in the real universe is $-3<n<0$ and so lies within
the regime where self-similar clustering may be a good approximation. 
There have now been quite extensive N-body tests of scaling behaviour for 
$-2<n<0$, and by and large self-similarity has been verified for the
statistics analysed so far (Efstathiou et al 1988; Lacey \& Cole
1994). The simulations become progressively more
challenging as $n$ becomes more negative, and the results for $n=-2$ are
significantly less convincing than those for larger $n$. Analysis of these
data show that most dark halos can be well represented as monolithic systems
with little substructure. Significant exceptions are almost always systems
in the process of merging or small objects which have fallen relatively
recently into a much more massive halo. Thus a good first description
of self-similar clustering is in terms of an abundance $A(M/M_*)dM/M_*^2$
of nonlinear ``dark halos'', a model for the internal structure of
these halos, and a rate 
$R(M_1/M_*,M_2/M_*)dM_1dM_2da/aM_*^4$ for mergers between halos. The abundance
and rate functions, $A$ and $R$, give us statistical information about the 
formation epochs, lifetimes, and evolution paths of dark halos, and 
surprisingly successful models for them are obtained from extensions of the 
Press \& Schechter argument. In contrast, the internal structure 
of individual halos can only be studied effectively through direct simulation.
I review these two approaches in the next few sections. Note that the 
extension from the self-similar 
case to more realistic initial conditions turns out to be straightforward. 
\medskip
\noindent{{\it 2.3 The P\&S model for clustering statistics~~~}The original 
derivation of a mass function for collapsed halos by Press \&
Schechter (1974)
was far from convincing, but several recent developments have given it a new 
lease of life. The first was the demonstration that an independent argument 
based on excursion set theory leads to an identical formula (Bond et
al 1991). The second was
the realisation that extensions of the argument allow the construction
of a much more complete but still relatively simple theory for hierarchical
clustering (Bower 1991; Bond et al 1991; Lacey \& Cole 1993). Finally, 
detailed comparisons with N-body simulations showed that 
the statistical predictions of the theory for mass functions, formation times, 
merger rates, etc. are in good (although clearly not perfect) agreement
with experiment (Lacey \& Cole 1994). A less comforting discovery is
the fact that the theory
works very poorly when its predictions are compared with simulation data on a 
halo by halo basis {\it i.e.} that the mass of the halo to which a given 
particle is predicted to belong by its excursion set trajectory is almost 
unrelated to the mass of the halo to which it actually belongs (White 1996).}

The P\&S formula for the probability that at time $t$ a random mass element 
is part of a halo with mass in the range $(M,dM)$ is
$$
f(M,t)dM={1\over\sqrt{2\pi}}{\sigma(M_*)\over\sigma(M)}{d\ln \sigma^2\over
d\ln M}\exp\left({-\sigma^2(M_*)\over 2\sigma^2(M)}\right){dM\over M}
$$
where $\sigma^2(M)$ is the initial (linear) variance on scale $M$ and
$M_*(t)$ is the characteristic nonlinear mass at time $t$ defined
by $b(t)\sigma(M_*)=\delta_c=1.686..$  In the excursion set derivation
$\sigma^2(M)$ is calculated as the total power in fourier modes with
wavenumber $k< k_c(M)=(6\pi^2\overline{\rho}a^3/M)^{1/3}$. Similarly, if we
consider a halo which has mass $M_2$ at time $t_2$, then according to the
extended theory the fraction of its material which was in halos of mass
in the range $(M_1,M_1+dM_1)$ at the earlier time $t_1$ (hence $t_1<t_2$
and $M_1<M_2$) was
$$
f(M_1,M_2,t_1,t_2)dM_1={1\over\sqrt{2\pi}}{\Delta\sigma_*\over\Delta\sigma}
{d\ln (\Delta\sigma)^2\over d\ln M_1}\exp\left({-(\Delta\sigma_*)^2\over 
2(\Delta\sigma)^2}\right){dM_1\over M_1}
$$
with $\Delta\sigma_* = \sigma(M_*(t_1)) - \sigma(M_*(t_2))$ and $(\Delta\sigma)^2
= \sigma^2(M_1) - \sigma^2(M_2)$. As Lacey, Cole, Bower, Kauffmann and others 
have shown these two formulae can be combined and used to derive merger rates,
distributions of formation and survival times, and merger histories which
agree well with those derived directly from numerical experiment. The point I
want to emphasise here concerns the structure of these equations rather than 
their precise form. The initial fluctuation spectrum enters only through
its variance $\sigma^2(M)=\sum_{k<k_c(M)}P(k)$ and time enters only through
the variance associated with the characteristic mass $M_*(t)$ and so through
the linear theory growth factor $b(t)$ which is used to define $M_*$.
When expressed in terms of these natural ``mass'' and ``time'' variables
the structure of hierarchical clustering is independent of the specific
cosmology under consideration, at least in the P\&S model. This is a 
tremendous simplification.

Another important simplification is the following. Let us consider a mass
element which is part of a halo of mass $M_2$ at time $t_2$ and part of
a halo of mass $M_1$ at time $t_1<t_2$. We can ask for the probability
that this element is part of a halo of mass $M_0<M_1$ at the yet earlier time 
$t_0$. In principle we might expect this probability to depend on $M_2$ and 
$t_2$ as well as on $M_1$ and $t_1$ but the excursion set derivation of the
P\&S theory shows that this is not the case. The probability is
just $f(M_0,M_1,t_0,t_1)dM_0$ as given by changing subscripts in the above 
formula. Thus the formation histories of the halos present at time $t_1$
do not depend on whether those halos are later incorporated into a more 
massive system. This shows that one must be careful when discussing how 
hierarchical clustering can introduce ``bias'' into the galaxy distribution.
According to P\&S theory a $10^{12}M_\odot$ halo at $z=1$ does not ``know''
whether it will be incorporated into a rich cluster or remain in a void at 
$z=0$. As a result the galaxy population contained in protocluster halos must 
be the same as that contained in protovoid halos of the same mass. Any bias in
galaxy population must arise either from the fact that the {\it distribution} 
of halo masses is different in protocluster and protovoid regions, or from 
the fact that the galaxies evolve in different environments between $z=1$ and 
the present. Both can plausibly lead to large systematic effects. It is unclear
to me whether this particular aspect of P\&S theory is realistic, since halos 
in N-body simulations clearly do know about their environment, at least to the 
extent that they often align with large-scale filaments.

An important property of hierarchical clustering which was first thoroughly
explored by Lacey \& Cole (1993) concerns the distribution of
formation times of
halos. They define the formation time of a halo to be the first time at
which its largest progenitor contains more than half the final mass,
and they show that the distribution of such formation times
depends weakly on the shape of the initial fluctuation spectrum but
strongly on halo mass. They find the typical formation time $t_f(M,t)$ for a 
halo of mass $M$ identified at time $t$ to be given by
$$
b(t)/b(t_f) -1 \sim \sigma(M)/\sigma(M_*(t)).
$$
For the particular case of scale-free clustering in
an Einstein-de Sitter universe they find the median redshift of formation
for halos of current mass $M$ to be
$$
z_f(M) = (2^{(n+3)/3}-1)^{1/2} (M/M_*(t_0))^{-(n+3)/6}.
$$
Fitting the abundance of rich clusters in the present universe to this kind
of hierarchical model implies that $M_*(t_0)\approx 2\times 10^{13}
\Omega_0^{-0.7}h^{-1}M_\odot$ so that clusters themselves are $20M_*$ 
events for $\Omega_0=1$ but only $6M_*$ events for $\Omega_0=0.2$. Thus
clusters are predicted to form very recently in an Einstein-de sitter universe
but less recently in a low density universe (the effect comes partly
from the reduction in $M/M_*$ and partly from the difference in the
behaviour of $b(t)$). A recent formation epoch seems to accord well with the
large fraction of real clusters which are observed to have significant
substructure and to be far from equilibrium, so this argument has been used 
to infer relatively large values of $\Omega_0$. Exactly how large a value is
required is still a matter of debate. In contrast, the halos of isolated 
galaxies have masses well below $M_*$ implying typical formation redshifts 
above unity for any $\Omega_0$. They are thus predicted to be well relaxed 
systems with a much lower incidence of substructure. Their last major
merging events are expected to be comfortably far in the past in most cases.
\medskip
\noindent{{\it 2.4 The faint galaxy problem~~~}For scale-free initial 
conditions the P\&S formula for the abundance of dark halos becomes
$$
N(M,t)dM=A(M/M_*){dM\over M_*^2}=\left({2\over\pi}\right)^
{1\over 2}{\overline{\rho}\over M_*}{n+1\over 3}\left(M\over 
M_*\right)^{n-9\over 6} \exp\left[- {1\over 2}\left({M\over M_*}\right)^
{3+n\over 3}\right]{dM\over M_*}.
$$
Thus a power-law, $N\propto M^{(n-9)/6}$, is truncated exponentially above
the characteristic mass $M_*(t)$. Recalling that the appropriate value for
$n$ is probably in the range $n\leq -1$ it is clear that the P\&S model
predicts that hierarchical clustering should give a very large number of
low mass halos in the present universe. For example, adopting $n=-1$ and the 
value of $M_*(t_0)$ quoted above, the predicted abundance of halos with masses
in the range $10^{10} < hM/M_\odot < 10^{11}$ is  $\sim 0.9 
\Omega_0^{1.23}h^3$ per cubic Mpc, and an even larger number is predicted 
for more negative $n$. For comparison, integrating the luminosity function
of Loveday and collaborators all the way down to $0.001L_*$ (this is well
below the effective limit of their observations) gives a total of only
$0.09h^3$ galaxies per cubic Mpc. This is a serious discrepancy, particularly
since many of the observed faint galaxies are actually satellites of brighter
systems or members of galaxy groups and so are contained in halos with masses
well above $10^{11}M_\odot$. Thus it seems that if $\Omega_0=1$ more than
90\% of all halos in the mass range $10^{10} < hM/M_\odot < 10^{11}$ must 
contain no galaxies of the kind represented in the catalogues used to compile
luminosity functions. Within hierarchical models it is certainly a challenge
to understand why this might be the case. The discrepancy is eliminated if
we are willing to accept $\Omega_0\sim 0.1$.

One resolution of this problem might be that the P\&S theory incorrectly
predicts the low mass end of the halo abundance distribution. There are a 
number of papers in the literature which discuss this possibility but
they come to no clear consensus. I believe that this is unlikely to be the 
answer, since high resolution N-body simulations have now been able to check 
the P\&S abundance against scale-free models with $0>n>-2$ and over the 
mass range $0.01 < M/M_* < 50$. The comparison is not straightforward since
the N-body mass functions depend on how ``halos'' are identified and the
P\&S functions depend on exactly how $M_*$ and $\sigma^2(M)$ are defined.
Nevertheless, the shape of the low mass tail is quite well fit in all cases,
and, if anything, the simulations seem to show slightly more mass in low
mass halos than is predicted by the theory. As $n$ becomes more negative
halos become less well separated from their environment, and for $n=-2$ 
many ``halos'' are poorly approximated as ellipsoidal equilibrium systems. 
This may be signalling a breakdown of the clustering hierarchy as $n$ 
approaches $-3$ and so might invalidate the P\&S abundance
predictions. Unfortunately this does not appear to 
solve the problem for models like CDM. From a high resolution simulation of 
standard CDM normalised to produce the correct rich cluster abundance I 
estimate $1.6h^3$ halos per cubic Mpc in the above mass range, well
above the observed galaxy abundance. A breakdown must occur in
models where $k^3P(k)$ reaches a maximum at some finite $k$ and thereafter
decreases. Such models are no longer ``hierarchical'';
they have a well defined initial coherence scale and they do not form
a significant number of objects below the corresponding mass. 
An example is the old Warm Dark Matter model, although I doubt that
this particular model is viable. 
\medskip
\noindent{{\it 2.5 Density structure of halos~~~} The first simulations of
the formation of dark halos in a CDM universe showed that they were predicted
to be monolithic ellipsoidal systems with a density structure that could be 
roughly approximated as ``isothermal'' {\it i.e.} $M(r)\propto r$
(Frenk et al 1988). Axis ratios
spanned the range between nearly prolate and nearly oblate, and values 
exceeding $2:1$ were quite common. More recent work with much better resolution
has confirmed these conclusions and shown that halo shapes remain far from
spherical even in their inner regions. This suggest some possible tests of the
theory. In disk galaxies deviations of the potential from axisymmetry can
be measured from the dynamics of polar rings or from the
photometric axis ratio of face-on systems. Results from the former test have
been mixed, but the latter one suggests that galaxy potentials are much more
nearly axisymetric than is predicted (Rix \& Zaritsky 1995).
 This test is not definitive since there
is a substantial contribution to the potential from the observed stars and gas,
and the accumulation of the galaxy could thus plausibly have modified the inner
structure of its halo. A similar test can be made using the X-ray emission
from galaxy clusters. This traces the potential at radii where the baryonic
contribution is thought to be small. Recent results show an ellipticity
distribution which is quite consistent with that expected for cluster mass
dark halos, but the interpretation is complicated by the abundant evidence 
for nonequilibrium structure in both real and simulated clusters
(Buote \& Canizares 1996).}

Even in lower mass halos where nonequilibrium effects are less important, 
high resolution simulations have shown that the isothermal density model is
a serious oversimplification. In the first place it is usually possible to
find a few small subclumps which have recently been accreted and have not
yet been disrupted by the main halo. More importantly, the density profiles
never have a constant logarithmic slope. Rather $\gamma=-d\ln\rho/d\ln r$ 
increases steadily with radius over the resolved region in almost all cases.
In simulations carried out to date there is no convincing evidence that
$\gamma$ is ever drops significantly below unity in the inner regions.
If a constant density ``core'' does form, it has yet to be resolved. This
has interesting consequences both for the rotation curves of dwarf galaxies,
and for the inner regions of galaxy clusters.

Ongoing work by a collaboration led by Julio Navarro is looking systematically
at the density profiles of dark halos in scale-free and CDM
universes with a variety of $\Omega_0$ values. The resolution 
limit of our simulations is
in all cases about 1\% of the outer radius of a halo (which we define as 
$r_{200}$, the radius at which the enclosed overdensity drops to 200). Over 
this radial range and for halos spanning about four orders of magnitude in
mass, we find that the radial density profiles can be fit quite well by the 
simple formula
$$
{\rho(r)\over \rho_{crit}} = {\delta_cr_s\over r(1+r/r_s)^2}.
$$
This model gives a density profile which bends gradually from $\gamma=1$ at
small radii to $\gamma=3$ at large radii. Less than 1\% of the halo mass
lies in the unresolved central regions. Notice that because of the 
definition of $r_{200}$, the parameters $c\equiv r_{200}/r_s$ (the
concentration parameter) and $\delta_c$ (the characteristic density in
units of the critical 
density) are not independent; this model is a one parameter fitting formula 
for halos of given mass. For all power spectra we find that $c$ decreases
(and hence $\delta_c$ decreases) with increasing halo mass. For a CDM universe
this decrease is from $c\sim 20$ at $M/M_*\sim 0.01$ to $c\sim 5$ at
$M/M_*\sim 100$. The increase is stronger for initial power spectra with more
positive $n$. The scatter about the relation is about 0.1 in $\log
c$. It is interesting that these trends can be interpreted purely as a
reflection of differing formation epoch. We find that for a suitable 
definition of formation redshift $z_f$ the relation
$\delta_c=1000\Omega_0(1+z_f)^3$ is a good description of our
numerical data for all power spectra and for all $\Omega_0$ values we have
tried so far.
 
These results have a number of interesting implications (Navarro et al
1996). For a CDM universe
the inner regions of rich clusters are sufficiently concentrated to account
for the observed giant arcs without violating constraints placed by
the observed distribution of X-ray gas. On the other hand, the centres of
dwarf galaxy halos are too concentrated to be consistent with the solid-body
rotation curves observed for a number of faint dark matter-dominated systems.
Something in the history of these systems must have altered the inner structure
of their halos if they are to be consistent with the model. A general result
is that the density profiles of galaxy halos are not predicted to be
scaled down versions of those of cluster halos. Instead the halos of
galaxies should be substantially more concentrated.
This concentration is presumably enhanced by the accumulation
of the galaxy itself. For bright galaxies like our own the maximum of
the circular velocity curve of the ``bare'' halo is predicted to lie
well outside the current optical radius of the galaxy, so that a rising
rotation curve would be predicted if the visible material were gravitationally
insignificant. 
\medskip
\noindent{{\it 2.6 Rotation of halos~~~} One result which has remained quite
stable since the earliest simulations of hierarchical clustering concerns
the distribution of the spin of dark halos as measured by the parameter
$\lambda=JE^{1/2}/GM^{5/2}$ where $J$, $E$ and $M$ are the magnitudes of 
halo angular momentum, binding energy, and mass respectively. The distribution
of $\lambda$ is found to be almost independent of $M$, of $P(k)$ and of
$\Omega_0$. It depends weakly on the way in which halos are identified in
the simulations. The median value is $\lambda_m\sim 0.05$ but the scatter is 
large with values ranging all the way from $<0.01$ to $>0.1$
(e.g. Cole \& Lacey 1996). The main factor
determining the value of $\lambda$ appears to be the morphology of halo 
formation; halos which form by mergers of similar sized clumps tend to have
relatively large angular momenta. The value of $\lambda$ also
correlates weakly with
central concentration in the sense that halos with large $\lambda$ 
tend to have small $c$ values for their mass.  

A parameter which is easier to interpret in terms of galaxy properties than
the traditional spin parameter is $\Lambda = H_0J/MV_c^2$ where the circular
velocity is calculated from $V_c^2=GM/r$ in the inner regions of a halo,
say where the mean enclosed density is 1000 times the critical density.
The distribution of $\Lambda$ also depends only weakly on $M$, $P(k)$ and 
$\Omega_0$. For scale-free initial conditions with $n=-1$ I find median
$\Lambda$'s of 0.003 for $\Omega_0=1$ and 0.002 for $\Omega_0=0.1$; again
the scatter spans more than an order of magnitude and halos with larger
$\Lambda$ tend to be less concentrated. An exponential disk with scale
length $r_d$ and constant rotation velocity $V_d$ has specific angular
momentum $2r_dV_d$. If we equate this to the specific angular momentum of
the halo $J/M$ then we find $H_0r_d=\Lambda V_c^2/2V_d$. Thus if $V_c\approx
V_d$ we find that a galaxy with $V_d=220$ km/s, for example, could 
contain a disk with $r_d$ of 2 or 3 $h^{-1}$kpc. This is indeed close to
the observed scale lengths of real disk galaxies with this rotation velocity.
Note, however, that there is little room for significant transfer of
angular momentum from disk material to the halo during disk formation, and 
that disks formed at high redshift would have to be significantly smaller (by
a factor of $(1+z)^{1.5}$ for $\Omega_0=1$). Thus there is only marginally enough
angular momentum available in hierarchical clustering to form disks as large 
as those observed today, and it is difficult to argue that damped Ly$\alpha$
systems at redshifts of 2 or 3 are collapsed disks which are systematically 
larger than those seen nearby. I will return to this problem later.
\medskip
\noindent{\bf 3 Including gas}
\smallskip
If we extend the above modelling to include a gas component in addition
to the dark matter then processes other than gravity can affect the gas
and new effects become important. The simplest case is that of a nonradiative 
gas without heating (other than shock heating), cooling or star formation. 
Such a gas is often referred to as ``adiabatic'' even though it is repeatedly 
shocked. Numerical experiments with a small gas fraction which is initially 
cold and distributed like the dark matter show that by and large the gas
density still parallels that of the dark matter at late times. This is 
not exactly true, however, because shocks cause the gas to move differently 
from the dark matter during the collapse and merging of nonlinear objects.  
This separation usually results in a transfer of energy and angular 
momentum from the dark matter to the gas, so that the gas ends up slightly 
less concentrated than the dark matter (Navarro \& White 1993). For scale-free 
initial conditions and $\Omega_0=1$ we would expect self-similar behaviour.
The numerical experiments needed to check this have not yet been carried out, 
but it seems that they will lead to halos in which the gas fraction declines 
steadily at smaller radii and higher densities.
\medskip
\noindent{{\it 3.1 The overcooling problem~~~}If radiative cooling is
allowed then the gas and dark matter distributions diverge much more
drastically. The typical density of nonlinear objects scales with
redshift as $(1+z)^3$ and so their gas cooling time approximately as
$(1+z)^{-3}$ (provided their virial temperature exceeds $10^4$K). On
the other hand dynamical times scale as $(1+z)^{-1.5}$. This
difference  means that although the bulk of the intergalactic
gas in present-day galaxy clusters is unable to cool, all the gas in 
nonlinear objects at $z>3$ can cool for a similar gas
fraction. When gas in a halo cools it sinks to the centre 
until collapse is stopped by rotation, by conversion into stars, or
by energy input of some kind. Simulations including cooling but no
star-formation or heating form small centrifugally supported disks,
whose apparent stability may well be an artifact of limited
numerical resolution (e.g. Navarro et al 1995). If collapse is 
stopped by rotation and star formation and there is no reheating, 
then a hierarchical model fails to make a realistic galaxy 
population. The problem is simply that all the gas is used up making
small objects at early times when cooling is efficient, so that
nothing is left to make big galaxies later on.

This overcooling problem has been known for 15 years. Two main ways of
circumventing it have been suggested. Star
formation in a small fraction of the gas in each halo may heat (and
perhaps eject) the rest, which is then available for incorporation
into later and larger objects. Alternatively, coupling the gas to an
ionizing background may prevent it from collapsing fully within
small potential wells, and thus from cooling in such objects. Although
both ideas seem feasible, further detailed
modelling is needed to show whether they work in practice. The few
simulations done so far show results which are dramatically dependent
on exactly how the additional physics is included (Navarro \& White
1993). Note that the
overcooling problem is less severe for CDM-like models than for the
scale-free models which Martin Rees and I originally worked out; 
the CDM power spectrum has so little small scale power that
even at quite late times a substantial fraction of the material is 
predicted to be in objects which have $t<10^4$K and so cannot cool.
This material can be incorporated into the large objects which form at
late times and so supply raw material for the formation of big
galaxies.
\medskip
\noindent{{\it 3.2 Angular momentum problems with gas~~~}I noted above
that hierarchical clustering produces barely enough angular
momentum to account for that observed in spiral disks. This becomes a 
serious problem in simulations of hierarchical cluatering which
include a cooling gas. As noted above most of the gas in such 
simulations settles to the centre of the small lumps present at early 
times. When these lumps merge to form the ``spiral halo'' their gas
cores also merge to form the ``spiral disk''. However, during this
merging the gas cores lose a large fraction (typically 80\%) of their 
orbital angular momentum to the dark matter, and as a result the disk
ends up much smaller than expected given the specific angular momentum
of its halo (Navarro et al 1995). The sizes predicted 
are well below those of observed
disks, so this particular formation path can be ruled out. The
problem is clearly that hierarchical clustering produces disks as
large as those observed only if two conditions are satisfied: (i)
disks must form late, probably after $z=1$, and (ii) they must form
from diffuse material rather from gas that has already condensed to
the centre of progenitor halos. It is the second condition that avoids
substantial transfer of angular momentum from gas to dark matter
during disk formation.
\medskip
\noindent{\bf Phenomenological models}
\medskip
Over the past five years there has been a substantial effort devoted
to developing phenomenological models for the formation of galaxies
in hierarchical clustering. Such models start from a description of
the clustering process based on P\&S theory or the ``peaks'' theory
of Bardeen and coworkers. This is combined with
simple models for the internal structure of nonlinear objects, for the
cooling of gas within them, for the conversion of that gas into stars,
for the feedback generated by star formation, and for the merging of
galaxies. Population synthesis models can be used to calculate colours
and luminosities for the galaxies, while chemical evolution models can
give their metallicities. I do not have enough space here to describe
these models in any detail but I think it is important to realise
their capabilities, and to recognise that they are currently much more
effective than numerical simulations for developing an understanding
of how the physical processes involved in galaxy formation shape the
observable properties of the galaxy population (see White \& Frenk 1991;
Kauffmann et al 1993,1994; Cole et al 1994; Kauffmann 1995; Heyl et al 1995).

Properties that can be relatively easily calculated by these
techniques include
the joint distributions of luminosity, colour, bulge-to-disk ratio,
metallicity, gas fraction, and halo circular velocity, together with 
the dependence of these distributions on  environment and redshift. 
These allow predictions for galaxy counts and redshift distributions as
a function of colour and morphology. One can also calculate age distributions 
for stars in the disks and bulges of galaxies for comparison with the
Galactic disk and bulge or with the properties of ellipticals in
different environments and at different redshifts. Furthermore the evolution 
of the galaxy populations in rich clusters can be analysed quite
easily. Additional levels of uncertainty are introduced if one
attempts to model properties which depend on galaxy size and there has
so far been little work on this area. Extensions of these methods
also allow the analytic treatment of issues related to
the spatial distribution of galaxies, for example ``bias'' as a
function of galaxy type and luminosity. Work on these problems is only
just beginning (e.g. Kauffmann et al 1996).

Results published so far show that many observed systematics which
were previously thought puzzling are natural consequences of these
hierarchical clustering models. Examples include the morphology
environment connection, the Butcher-Oemler effect, the Tully-Fisher
relation and its small scatter, the fact that bulges look like
ellipticals yet lie within disks which could not be merger products,
the fact that rich clusters contain old ellipticals even near
$z=1$, the simultaneous observation of steep counts and
``no-evolution'' redshift distributions for faint galaxies. There are
also a few serious problems, the worst being an overabundance of faint
galaxies which is a direct consequence of the halo overabundance problem
noted above. Some progress has already been made towards reducing
this discrepancy, and at the moment the hierarchical
clustering picture seems to provide a remarkably good description of
the observed galaxy populations and their evolution with redshift.
\bigskip
\centerline{\bf REFERENCES}
\medskip

Barnes, J. 1995. Interactions and Mergers in Galaxy Formation. In: The
Formation and Evolution of Galaxies, ed. C. Munoz-Tunon and
F. Sanchez, pp 399-454,  Cambridge: Cambridge Univ. Press.

Bond, J.R., S.M. Cole, G. Efstathiou and N. Kaiser. 1991 Excursion Set
Mass Functions for Hierarchical Gaussian Fluctuations. {\it ApJ}
{\bf 379}, 440-463.

Buote, D.A. and C.R. Canizares. 1996. X-ray Constraints on the
Intrinsic Shapes and Baryon Fractions of Five Abell Clusters.
{\it ApJ}, in press.
 
Bower, R.J. 1991. Evolution of Groups of Galaxies in the
Press-Schechter Formalism. {\it MNRAS} {\bf 248}, 332-352.

Cole, S.M., A. Aragon-Salamanca, C.S. Frenk, J.F. Navarro and
S.E. Zepf. 1994. A recipe for Galaxy Formation.
{\it MNRAS} {\bf 271}, 781-798.

Efstathiou, G., C.S. Frenk, S.D.M. White and
M. Davis. 1988. Gravitational Clustering from Scale-free Initial
Conditions.  {\it MNRAS} {\bf 235}, 715-748.

Frenk, C.S., S.D.M. White, M. Davis and G. Efstathiou. 1988. Formation
of Dark Halos in a Universe dominated by Cold Dark Matter.
{\it ApJ} {\bf 327}, 507-525.

Hausman, M. and J.P. Ostriker. 1978. Galactic Cannabalism III.
{\it ApJ} {\bf 224}, 320-336.

Heyl, J.S., S.M. Cole, C.S. Frenk and J.F. Navarro. 1995. Galaxy
Formation in a Variety of Hierarchical Models.
{\it MNRAS} {\bf 274}, 755-770.

Kauffmann, G., B. Guiderdoni and S.D.M. White. 1994. Faint Galaxy
Counts in a Hierarchical Universe.
{\it MNRAS} {\bf 267}, 981-999.

Kauffmann, G., S.D.M. White and B. Guiderdoni. 1993. The Formation and
Evolution of Galaxies within Merging Dark Matter Halos.
{\it MNRAS} {\bf 264}, 201 218.

Kauffmann, G. 1995. The Properties of High Redshift Cluster Galaxies.
{\it MNRAS} {\bf 274}, 161-172.

Kauffmann, G., A. Nusser and M. Steinmetz. 1996. Galaxy Formation and
Large-scale Bias. {\it MNRAS}, in press.

Lacey, C.G. and S.M. Cole. 1993. Merger rates in Hierarchical Models
of Galaxy Formation I. {\it MNRAS} {\bf 262}, 627-641.

Lacey, C.G. and S.M. Cole. 1994. Merger rates in Hierarchical Models
of Galaxy Formation II. {\it MNRAS} {\bf 271}, 676-692.

Navarro, J.F. and S.D.M. White. 1993.Simulations of Dissipative Galaxy
Formation in hierarchically Clustering Universes - I.
{\it MNRAS} {\bf 265}, 271-300.

Navarro, J.F., C.S. Frenk and S.D.M. White. 1995. Assembly of Galaxies
in a Hierarchically Clustering Universe. 
{\it MNRAS} {\bf 275}, 56-66.

Navarro, J.F., C.S. Frenk and S.D.M. White. 1996. The Structure of
Cold Dark Matter Halos. {\it ApJ}, in press.

Padmanhaban, T. 1993. Structure Formation in the Universe. Cambridge:
Cambridge Univ. Press.

Peebles, P.J.E. 1978. Stability of a Hierarchical Pattern in the
Distribution of Galaxies. {\it A{\&}A} {\bf 68}, 345-352.

Peebles, P.J.E. 1980. The Large-scale Structure of the
Universe. Princeton: Princeton Univ. Press.

Peebles, P.J.E. 1993. Principles of Physical Cosmology. Princeton: 
Princeton Univ. Press.

Press, W.H. and P.L. Schechter. 1974. Formation of Galaxies and Galaxy
Clusters by Self-similar Gravitational Condensation. {\it ApJ} 
{\bf 187}, 425-438.

Rix, H.-W. and D.F. Zaritsky. 1995. Non-axisymmetric Structures in the
Stellar Disks of Galaxies. {\it ApJ} {\bf 447}, 82-91. 

Toomre, A. 1977. Mergers and some Consequences. In: Evolution of
Galaxies and Stellar Populations, eds B.M. Tinsley and R.B. Larson,
pp 401-417, New Haven: Yale Univ Observ.

Toth, G. and J.P. Ostriker. 1992. Galactic Disks, Infall and the
Global Value of $\Omega_0$. {\it ApJ} {\bf 389}, 5-26.

White, S.D.M. and M.J. Rees. 1978. Core Condensation in Heavy Halos: a
Two Stage Theory for Galaxy Formation and Clustering.
{\it MNRAS} {\bf 183}, 341-358.

White, S.D.M., C.S. Frenk and M. Davis. 1983. Clustering in a
Neutrino-dominated Universe. {\it ApJ} {\bf 274}, L1-L5.

White, S.D.M., C.S. Frenk, M. Davis and G. Efstathiou. 1987. Clusters,
Filaments and Voids in a Universe dominated by Cold Dark Matter.
{\it ApJ} {\bf 313}, 505-516.

White, S.D.M. and C.S. Frenk. 1991. Galaxy Formation through
Hierarchical Clustering. {\it ApJ} {\bf 379}, 52-79.

White, S.D.M. 1996. Formation and Evolution of Galaxies, In: Cosmology
and Large-scale Structure, eds R. Schaeffer, J. Silk and
J. Zinn-Justin, in press, Dordrecht: Elsevier Science.

\end